\documentstyle[preprint,floats,prd,aps]{revtex}
\input epsf.tex
\def\DESepsf(#1 width #2){\epsfxsize=#2 \epsfbox{#1}}
\begin{document}
\preprint{\vbox{\hbox{OITS-564}}  }
\draft
\title{Long Distance Contributions to Penguin Processes \\
$b\rightarrow s\gamma$
and $b\rightarrow d \gamma$
\footnote{Work supported in part by the Department of Energy Grant No.
DE-FG06-85ER40224.}}
\author{N.G. Deshpande, Xiao-Gang He, and Josip Trampetic\footnote{On leave of
absence from the Dept. of Theor. Phys., R. Boskovic Inst., Zagreb,
Croatia.}}\address{Institute of Theoretical Science\\
University of Oregon\\
Eugene, OR 97403-5203, USA}

\date{Revised January 23, 1995}
\maketitle
\begin{abstract}
We consider the long distance contributions to inclusive
penguin processes through processes like $b\rightarrow s V$ and $b\rightarrow
d V$ where $V$ are $^3S_1(c\bar c)$ states $\psi_i$ in the former case and
include $\rho$, $\omega$ for the latter case. We carefully examine
vector dominance for $\bar c c$ states, and conclude that there is a large
suppression of $\psi\sim \gamma$ transition when $\psi$ is at $q^2 = 0$.
The long distance effects can be at most 10\% in
both the amplitudes for $b\rightarrow s\gamma$ and
$b\rightarrow d \gamma$. Although the long distance contributions are small,
the ratio $BR(b\rightarrow d\gamma)/BR(b \rightarrow s \gamma) =
|V_{td}/V_{ts}|^2$ does not hold due to significant $u$ and $c$ loop
contributions to the short distance $b\rightarrow d\gamma$ amplitudes .
\end{abstract}
Penguin dominated process $b\rightarrow s \gamma$ and the hadronic mode
$B\rightarrow K^* \gamma$ have both been measured recently at CLEO\cite{cleo}.
Since these processes arise through loop diagrams, it has been stressed from
the inception that they provide important tests of the Standard Model (SM). The
theoretical calculation of the fundamental process at the quark level has been
improved over successive years so that a complete perturbative QCD calculation
to leading order involving up to two loop diagrams now
exists\cite{desh,gamma,bsg}. The next to leading order calculation is only
partially completed\cite{nlt}. Similarly, the exclusive mode $B\rightarrow
K^*\gamma$ has received increasing attention, and form factors have been
calculated\cite{desh1} in various quark models; using QCD sum rules; using
effective heavy quark theory combined with chiral perturbation theory; and
using QCD lattice gauge techniques. These calculations are in agreement with
the data within the limits of both experimental and theoretical uncertainties.
A large number of papers have also been written setting limits on physics
beyond the SM which could affect $ b\rightarrow s \gamma$ process. For a recent
survey the review of Hewett summarizes the situation\cite{hewett}.

Importance of these penguin process leads us to examine another type of
contribution to the quark processes $b\rightarrow s/d +\gamma$ that
has hitherto been neglected. At the quark level, the process $b\rightarrow
s/d+V$ where $V$ are vector mesons, typically $\psi$ and
its radial excitations, and $\rho$, $\omega$, is a good description of
the inclusive $V$ production. These vector bosons can be
converted to a photon from the usual vector dominance arguments. Although
vector
dominance is known to work quite well for $\rho$ and $\omega$, the situation
with $\psi_i$ is not clear because it involves extrapolation from $q^2\approx
9$GeV$^2$
to $q^2 = 0$. We carry out a critical analysis and find a suppression factor in
this case.
We note in passing that such resonance effects have been considered previously
in the context of $B\rightarrow X_s l^+l^-$ which proceeds at the quark level
through $b\rightarrow s l^+l^-$\cite{res}. The dilepton mass spectrum obviously
has peaks corresponding to the production of $\psi_i$, and the spectrum has to
be measured away from these peaks where resonance effects are small to get at
the short distance part. In the present problem, however, there is no
experimental way of removing contributions due to vector bosons in the
intermediate states.
The importance of long distance contribution for the exclusive modes like
$B\rightarrow K^*\gamma$ and $B\rightarrow \rho \gamma$
has been noted by several authors recently\cite{long,long1,soni}. These
contributions at the quark level
subprocess has not been realized up to now.

The calculation of long distance contribution to $b\rightarrow s/d + \gamma$
that we have performed may also be thought of in a different way. For example,
the inclusive process $B\rightarrow X_s \gamma$ consists of a sum over large
number of exclusive channels where $X_s = K^*,\;K\pi,\;K\pi\pi$, etc. Each
channel has contribution from the pure penguin process and from long distance
contribution involving $\psi_i$ conversion to $\gamma$. We take into
account all these separate channels on the average by calculating
the inclusive process at the quark level.

First, we consider the process $b\rightarrow s\gamma$ with the additional
contribution coming from $b\rightarrow s\psi_i$, where $\psi_i$ are the set of
all $J=1$, $ l= 0$ ($c\bar c$) bound states, with $\psi_i$ to $\gamma$
conversion. The treatment for $b\rightarrow d\gamma$ is similar and will be
discussed later.
The leading order QCD corrected effective Hamiltonian describing the processes
is\cite{gamma}
\begin{eqnarray}
H_{\Delta B=1} &=& {G_F\over \sqrt{2}}[V_{cb}V^*_{cs}(c_1O_1 + c_2 O_2) -
V_{tb}V^*_{ts}c_7^{eff}O_7] +H.C.\;,
\end{eqnarray}
where $O_i$ are defined as
\begin{eqnarray}
O_1 &=& \bar s_\alpha \gamma_\mu(1-\gamma_5)c_\beta\bar
c_\beta\gamma^\mu(1-\gamma_5)b_\alpha\;,\nonumber\\
O_2 &=&\bar s \gamma_\mu(1-\gamma_5)c\bar
c\gamma^\mu(1-\gamma_5)b\;,\nonumber\\
O_7 &=& {e\over 8\pi^2}m_b\bar s \sigma^{\mu\nu}(1+\gamma_5)bF_{\mu\nu}\;.
\end{eqnarray}
The Wilson coefficients $c_i$ are given by
\begin{eqnarray}
c_1 &=& {1\over 2}(\eta^{6/23} - \eta^{-12/23})\;,\;\;\;\;
c_2 = {1\over 2}(\eta^{6/23} + \eta^{-12/23})\;,\nonumber\\
c_7^{eff} &=& \eta^{16/23}(c_7(x) -{8\over 3}(1-\eta^{-2/23})c_8(x))
+\sum_i^8h_i\eta^{p_i}\;,\nonumber\\
c_7(x) &=&{x(7-5x-8x^2)\over 24(x-1)^3} - {x^2(2-3x)\over
4(x-1)^4}\mbox{ln}x\;,\nonumber\\
c_8(x) &=& {x(2+5x-x^2)\over 8(x-1)^3} -{3x^2\over 4(x-1)^4} \mbox{ln}x\;,
\end{eqnarray}
where $\eta = \alpha_s(m_W)/\alpha_s(\mu)$, $x=m^2_t/m_W^2$. We use $h_i$ and
$p_i$ calculated in \cite{bsg}
\begin{eqnarray}
(h_1,& h_2&, h_3, h_4, h_5, h_6, h_7, h_8)\nonumber\\
&=&(2.2996, -1.0880, -0.4286, -0.0714,-0.6494, -0.0380, -0.0186,
-0.0057)\;,\nonumber\\
(p_1,& p_2&, p_3, p_4, p_5, p_6, p_7, p_8)\nonumber\\
&=&(0.6087, 0.6957, 0.2609, -0.5217, 0.4086, -0.4230, -0.8994, 0.1456)\;.
\end{eqnarray}

In the above Hamiltonian we have neglected terms arising from gluon and
electroweak penguins since their contributions are found to be small. The
short distance amplitude $M_{SD}$ for $b\rightarrow s \gamma$ is given by the
term involving $O_7$.

Let us now discuss the process $b\rightarrow s\psi$.
Using factorization, we obtain the inclusive decay amplitude for
$b\rightarrow s \psi$,
\begin{eqnarray}
M(b\rightarrow s \psi(q)) = ig_\psi(q^2) {G_F\over \sqrt{2}} a_2 V_{cb}V_{cs}^*
\bar s \gamma^\mu (1-\gamma_5)b \epsilon^\dagger_\mu(q)\;,
\end{eqnarray}
where $g_\psi$ is defined as $<\psi(q)|\bar c\gamma_\mu c|0> = ig_\psi(q^2)
\epsilon^\dagger_\mu(q)$, and $a_2 = c_1 +c_2/N$ with $N$ being the number of
colors. It is well known that factorization approximation works only if one
treats the constants $a_1 = c_1/N+c_2$ and $a_2= c_1+c_2/N$ as phenomenological
parameters. Using experimental data\cite{pdg} on total inclusive $\psi$
production $BR(B\rightarrow X_s \psi) = (1.31\pm 0.17)\%$ and removing $\psi$
production from cascade processes $B\rightarrow X_s (\psi(2S)\;, \;
\chi_{c1}(1P))$ with $(\psi(2S)\;, \; \chi_{c1}(1P)) \rightarrow \psi$, we find
the branching ratio for direct $B\rightarrow  X_s \psi$ to be $
(0.74 \pm 0.23)\%$. From this and the semileptonic partial width we deduce the
value of $|a_2| = 0.24\pm0.04$. Further, from $B\rightarrow D\pi$ and
$B\rightarrow D\rho$, the sign of $a_2$ is determined
to be positive\cite{a2,L/T}. We shall assume this sign holds also for
$B\rightarrow
X_s \psi$.
We now use Vector Meson Dominance (VMD) to calculate the amplitude
for $b\rightarrow s\gamma$ from $b\rightarrow s\psi$ with the conversion of
$\psi$ to $\gamma$.
Note that the above amplitude in eq.(5) contains both transverse and
longitudinal $\psi$ polarizations. Gauge invariance requires that only the
transverse component of $\psi$ be converted to $\gamma$.
To separate this part of the amplitude, we decompose
$\bar s \gamma^\mu (1-\gamma_5) b$, using Gorden identity and neglecting $m_s$:
\begin{eqnarray}
m_b\bar s \gamma^\mu(1-\gamma_5) b
= (2P^\mu -q^\mu) \bar s (1+\gamma_5)b - i\bar s \sigma^{\mu\nu}q_\nu
(1+\gamma_5)b \;,
\end{eqnarray}
where $P^\mu$ is the b quark momentum and $q^\mu$ is the $\psi$ momentum.
Since only
terms with $\sigma^{\mu\nu}$ couple to transverse $\psi$, we obtain the
following transverse amplitude:
\begin{eqnarray}
M(b\rightarrow s\psi )_T
=  {G_F\over \sqrt{2}}a_2V_{cb}V_{cs}^* {g_\psi(m_\psi^2)\over m_b}
\bar s \sigma^{\mu\nu}(1+\gamma_5)b q_\nu \epsilon^\dagger_\mu(q)\;.
\end{eqnarray}

One may ask how good is such a model as a description of $B\rightarrow X_s
\psi$ process. Using the model we find the ratio of transverse to
longitudinal decay rates to be $\Gamma_T/\Gamma_L = 0.8$, implying
$\Gamma_L/\Gamma = 0.56$. The process $B\rightarrow X_s \psi$ unlike the quark
subprocess $b\rightarrow s \psi$ has a range of $\psi$ momenta corresponding to
$X_s$ invariant masses. Experimental data shows that for all events with
$p_\psi < 2 \mbox{GeV/c}$, $\Gamma_L/\Gamma = 0.59\pm 0.15$ are in good
agreement
with the model. If, however, one considers the highest momentum bin with $1.4
\mbox{GeV/c} < p_\psi < 2.0 \mbox{GeV/c}$, the ratio $\Gamma_L/\Gamma = 0.78\pm
0.17$\cite{L/T}. This momentum bin is known to be dominated by two body modes
where $X_s = K^*$ or K. The polarization of $\psi$ in exclusive modes depends
on details of hadronic form factors and it is inappropriate to use the
prediction of the inclusive model for $b\rightarrow s\psi$ at the extreme
values of $p_\psi$ where phase space constrains the final states. We shall
therefore employ this model as a reasonable description of the inclusive
process. One will have to bear in mind, however, that in the measurement of
$B\rightarrow X_s\gamma$, the gamma spectrum might be distorted at high
momentum precisely because two body modes might dominate this region. Thus in
extracting the rate for $b\rightarrow s\gamma$ from the measurement of
$B\rightarrow X_s\gamma$ care will have to be exercised in sampling a broad
enough region of gamma momenta.

Using the $\psi$ to $\gamma$ conversion vertex, $<0|eJ^\mu_{em}|\psi>
=(2e/3)<0|\bar c\gamma^\mu c|\psi> = -(2e/3)ig_\psi(0)\epsilon_\mu$, we have
for the long distance amplitude:
\begin{eqnarray}
M_{LD}(b\rightarrow s\psi\rightarrow s \gamma )
=  {G_F\over 2\sqrt{2}}a_2V_{tb}V_{ts}^* ({2\over 3}e{g_\psi^2(0)\over m_\psi^2
m_b})
\bar s \sigma^{\mu\nu}(1+\gamma_5)b F_{\mu\nu}\;.
\end{eqnarray}
Here we have used $V_{cb}V_{cs}^* = -V_{tb}V_{ts}^* - V_{ub}V_{us}^*$, and
neglected a small term proportional to $V_{ub}V_{us}^*$. Note that it is
necessary to
extrapolate the momentum of
$\psi$ from $q^2=m_\psi^2$ to $q^2 = 0$ in the $\psi$ to $\gamma$ conversion,
and we must use the value $g_\psi(0)$ for the conversion coupling constant.
Including all the ($c\bar c$) resonances and the short distance contribution
$M_{SD}$, we find the full amplitude to be
\begin{eqnarray}
M(b\rightarrow s\gamma)
=  -{eG_F\over 2\sqrt{2}}V_{tb}V_{ts}^*[{1\over 4\pi^2}m_b c^{eff}_7(\mu)
-a_2{2\over 3} \sum_i{g_{\psi_i}^2(0)\over m_{\psi_i}^2m_b} ]
\bar s \sigma^{\mu\nu}(1+\gamma_5)b F_{\mu\nu}\;,
\end{eqnarray}
where $\psi_i$ represents the following  vector $c\bar c$ resonant states:
$\psi(1S)$, $\psi(2S)$, $\psi(3770)$, $\psi(4040)$, $\psi(4160)$, and
$\psi(4415)$. Note that the relative sign between the long distance and short
distance contributions is not arbitrary but determined by the theory. The
various decay constants $g^2_{\psi_i}$ at $q^2 = m_{\psi_i}^2$ are calculated
using $g^2_{\psi_i} = 27 \Gamma(\psi_i \rightarrow e^+e^-)m_{\psi_i}^3/16\pi
\alpha^2$, to be
\begin{eqnarray}
&g^2_{\psi(1S)}&:g^2_{\psi(2S)}:g^2_{\psi(3770)}:g^2_{\psi(4040)}:g^2_{\psi(4160)}:g^2_{\psi(4415)} \nonumber\\
&=&1.575:1.080:0.140:0.499:0.559:0.408\;.
\end{eqnarray}
Here the units for $g^2_{\psi_i}$ are in GeV$^4$.

Let us now examine critically the use of VMD for $\psi_i$ to $\gamma$
conversion.
We use two different sets of experimental data together with VMD to estimate
$g_{\psi}$ at $q^2 = 0$: i) photoproduction of $\psi$, and ii) branching ratio
for $\chi_{c0}(1P)\rightarrow \gamma \gamma$ based on VMD from
$\chi_{c0}(1P)\rightarrow \gamma \psi$ and $\psi(2S) \rightarrow
\chi_{c0}(1P)\gamma$
decays.

i) In the case for photoproduction of $\psi$, the total $\psi N$
cross section $\sigma_T(\psi N)$ is related to the differential cross section
$d\sigma (\gamma N)/dt|_{t=0}$ for $\gamma N \rightarrow \psi N$ at $t = 0$ via
VMD, by\cite{13}
\begin{eqnarray}
{d\sigma(\gamma N)\over dt}|_{t=0} = \left ( {2eg_{\psi}(0)\over
3m^2_{\psi(1S)}}\right )^2{1\over 16\pi} \sigma_T(\psi N)^2\;.
\end{eqnarray}
The experimental data for $\sigma(\psi N) = 3.5\pm 0.5$ mb\cite{14}, and
$d\sigma (\gamma N)/dt|_{t=0}
= (52 \pm 11)$ nb/GeV$^2$\cite{15}, gives $\kappa
=g_{\psi(1S)}^2(0)/g_{\psi(1S)}^2(m_\psi^2)
= 0.12\pm 0.04$. This suppresses the long distance contribution to
$b\rightarrow s \gamma$. Unfortunately we do not have data on photoproduction
of other excited
$\psi$ particles. However, since $q^2$ extrapolation is of similar magnitude,
we
assume this suppression factor is universal.

ii) Within the VMD, the leading contributions to  $\chi_{c0}(1P) \rightarrow
\gamma\gamma$ are from $\chi_{c0}(1P) \rightarrow \psi(1S, 2S) \gamma$ with the
conversion of $\psi(1S, 2S)$ to $\gamma$. Since $\psi(2S)$ is heavier than
$\chi_{c0}(1P)$, we extract the relevant contribution from $\psi(2S)
\rightarrow \chi_{c0}(1P) \gamma$ decay, and find
\newpage
\begin{eqnarray}
&BR&(\chi_{c0}(1P) \rightarrow
\gamma\gamma) = \left ( {2e g_{\psi(1S)}(0)\over 3 m_\psi^2}\right )^2m_\chi^6
 {BR(\chi_{c0}(1P)\rightarrow \psi(1S) \gamma)\over
(m_\chi^2-m_\psi^2)^3}\nonumber\\
&\times&\left (1+ R{g_{\psi(2S)}(0)m_{\psi(1S)}^2\over g_{\psi(1S)}(0)
m_{\psi(2S)}^2}
\sqrt{
{\Gamma(\psi(2S)\rightarrow \chi_{c0}(1P) \gamma)\over
\Gamma(\chi_{c0}(1P)\rightarrow \psi(1S) \gamma)}}
 \right )^2\;,
\end{eqnarray}
where $m_\chi$ is the mass of $\chi_{c0}(1P)$ and
\begin{eqnarray}
 R=\sqrt{m_{\psi(2S)}^5(m_\chi^2-m_{\psi(1S)}^2)^3\over
m^3_{\chi} (2m_{\psi(2S)}^2-m_\chi^2)(m_{\psi(2S)}^2 -m_\chi^2)^3}\;.
\end{eqnarray}
The measured branching ratios
$BR(\chi_{c0(1P)} \rightarrow \psi(1S) \gamma) = (6.6\pm 1.8)\times 10^{-3}$,
$BR(\psi(2S) \rightarrow \chi_{c0}(1P) \gamma = (9.3\pm0.8)\%$, and
$BR(\chi_{c0}(1P)\rightarrow \gamma\gamma) = (4.0\pm 2.3)\times
10^{-4}$\cite{pdg}, then give the suppression factor $\kappa =
0.25\pm 0.16$.
Applying weighted average to the photoproduction  and decay data, we find
$\kappa = 0.13 \pm 0.04$.   This value is consistent with a  calculation based
on dispersion relation\cite{16}, and we shall adopt it here.

{}From the above discussions, we see that there is a large suppression for
$g_{\psi_i}$ when extrapolating from $q^2 = m_{\psi_i}^2$ to $q^2=0$. Using
the central value for $\kappa=0.13$, we find that the long distance
contribution
to $b\rightarrow s \gamma$ amplitude is about 3\% to 4\% of the short distance
contribution when  $a_2$ is varied between $0.20$ to $0.28$ with $m_b = 4.9$
GeV,
$m_c = 1.5$ GeV, $\Lambda_4 = 0.25$ GeV, and $\mu = m_b$. The short distance
amplitude depends weakly on the top quark mass varying between 150 GeV to 200
GeV. There
are uncertainties\cite{buras} in the short distance contributions. One of
them is the renormalization scale which is
not very well defined. If one chooses $\mu$ around $2m_b$, then the branching
ratio goes down by about 30\%. There are also indications that the
next-to-leading
order QCD correction may reduce the branching ratio\cite{nlt}. Including all
these uncertainties and the uncertainties in $a_2$ and the suppression factor
$\kappa$,
the long distance contribution can be as large as
10\% in amplitude of the short distance contribution. In Fig. 1. we plot the
rate for $b\rightarrow s \gamma$ from pure short distance contribution and
including long distance contribution. We have allowed $\mu$ to vary from $2m_b$
to $m_b/2$ and maximized long distance contribution by taking $a_2$ and
$\kappa$
values at the top of their $2\sigma$ range. The allowed range for the SM is
from the top solid line 1 to almost the lower dashed line 2. The experimental
limits
at 90\% C.L. are $1\times 10^{-4} < BR(b\rightarrow s\gamma) < 4\times 10^{-4}$
\cite{cleo}. The experiment clearly favors larger $\mu$ values. Complete next
to leading order calculation should reduce theoretical uncertainty related to
$\mu$ dependence, and improvements in experimental precision on $b\rightarrow
s\gamma$ will help to test the SM with higher accuracy.

The same analysis can be extended to for $b\rightarrow d \gamma$.
The short distance top quark loop contribution for $b\rightarrow d \gamma$
is obtained by replacing $s\rightarrow d$ in eq.(1). In this case, however,
there are significant contributions from $u$ and $c$ loops which introduce
terms proportional to $V_{ub}V_{ud}^*$ for the short distance amplitude. These
two loop contributions have log dependence on $u$ and $c$ masses\cite{17}.
Using the matching constant
determined in Ref.\cite{18}, we have to the lowest order in $\alpha_s$
\begin{eqnarray}
H = -{eG_F\over 8\pi^2\sqrt{2}}V_{ub}V_{ud}^*m_b {52\alpha_s(\mu)\over
81\pi}\mbox{ln}{m_c^2\over m_u^2}
\bar d \sigma^{\mu\nu}(1+\gamma_5)b F_{\mu\nu}\;.
\end{eqnarray}

To obtain long distance
amplitude in this case, one needs to consider
 more competing resonant states. To the leading order these resonant states
include: the $\psi_i$ states considered before, and $\rho$ and $\omega$ states.
We
obtain the long distance amplitude:
\begin{eqnarray}
M_{LD}(b\rightarrow d V\rightarrow d \gamma )
&=& -{eG_F\over 2\sqrt{2}}a_2[V_{cb}V_{cd}^* {2\over
3}\sum_i{g_{\psi_i}^2(0)\over
m_{\psi_i}^2 m_b}\nonumber\\
& +& V_{ub}V_{ud}^* ({1\over 2}{g_\rho^2(0)\over m_\rho^2 m_b} +{1\over 6}
{g_\omega^2(0)
\over m_{\omega}^2 m_b})]
\bar d \sigma^{\mu\nu}(1+\gamma_5)b F_{\mu\nu}\;,
\end{eqnarray}
where $|g_\rho(m_\rho^2)| = 0.166\mbox{GeV}^2$, and
$|g_\omega(m_\omega^2)|=0.152 \mbox{GeV}^2$ are
defined by $<\rho^0|\bar u\gamma_\mu u|0> = ig_\rho \epsilon^\dagger_\mu(\rho)
/\sqrt{2}$, and $<\omega^0|\bar u\gamma_\mu u|0> = ig_\omega
\epsilon^\dagger_\mu(\omega)/\sqrt{2}$. The conversion coupling constants
$g_\rho(0)$ and $g_\omega(0)$ do not change very much\cite{13,16} when
extrapolated from their values at $q^2 = m_{\rho,\omega}^2$. We shall use
$|g_\rho(0)| = 0.166\mbox{GeV}^2$, and $|g_\omega(0)|=0.152 \mbox{GeV}^2$.
 Adding the short distance contributions and using, $V_{cb}V_{cd}^* =
-V_{tb}V_{td}^* -V_{ub}V_{ud}^*$,
we have the following total amplitude
\begin{eqnarray}
&M&(b\rightarrow d \gamma )
= -{eG_F\over 2\sqrt{2}}[V_{tb}V_{td}^* ( {m_b\over 4\pi^2} c^{eff}_7(\mu)
-{2\over 3}a_2\sum_i{g_{\psi_i}^2(0)\over m_{\psi_i}^2 m_b} )\nonumber\\
&+& V_{ub}V_{ud}^*({m_b\over 4\pi^2}{52\alpha_s(\mu)\over
81\pi}\mbox{ln}{m_c^2\over m_u^2}- a_2 ( {2\over 3}\sum_i{g_{\psi_i}^2(0)\over
m_{\psi_i}^2
m_b}-{1\over 2}{g_\rho^2(0)\over m_\rho^2 m_b} -{1\over 6} {g_\omega^2(0)
\over m_{\omega}^2 m_b}))]\nonumber\\
&\times&
\bar d \sigma^{\mu\nu}(1+\gamma_5)b F_{\mu\nu}\;.
\end{eqnarray}
The long distance contributions are all found to be small, relative to the
short
ones. Also, the contributions
from $\psi_i$ cancel those from $\rho$ and $\omega$ for the
term proportional to $V_{ub}V_{ud}^*$. For the central values for $a_2$ and
$\kappa$, this cancellation is almost complete. The long distance contribution
to the term proportional to $V_{tb}V_{td}^*$ is about 3.5\% of the top loop
short distance contribution. Varying $\mu$ from $2m_b$ to $m_b/2$ and
$a_2$ and $\kappa$ in their $2\sigma$ ranges, we find that the long distance
contribuitons to the term proportional to $V_{ub}V_{ud}^*$ is less than 3\%.
Using the dynamic u quark mass of 300 MeV, we find the ratio of the u and c
loop short distance contributions to the top loop one to be
$-0.4V_{ub}V_{ud}^*/V_{tb}V_{td}$. It is clear that even though the long
distance
contributions do not badly destroy the relation $BR(b\rightarrow d\gamma)
/BR(b\rightarrow s\gamma) = |V_{td}/V_{ts}|^2$, the short distance
contributions
from u and c loop do.

Similar analysis can be carried out for exclusive radiative $B$ decays. It has
been pointed out that the long distance contribution to the $B\rightarrow
K^*\gamma$ amplitude
is small (less than 10\%) using a value $\kappa \approx 0.4$\cite{long1}. Using
the value
obtained by us, the contribution would be even smaller. The
long distance
contributions to $\bar B^0\rightarrow \rho^0 \gamma$ and $\bar B^0\rightarrow
\omega \gamma$
amplitudes are also small. The long distance VMD contributions are all
proportional to $a_2$ in $B\rightarrow K^*\;(\rho^0,\;\omega)\gamma$. However,
for the charged mode
$B^-\rightarrow \rho^-\gamma$ there is a crucial difference because of the
presence of a term proportional to $a_1$, which is about 4 times larger than
$a_2$ in magnitude. This term originates from an annlihilation diagram,
and it does not contribute to the spectator inclusive $b\rightarrow d\gamma$
decay discussed above. One should carefully consider all possible long distance
contributions to the exclusive modes.
We will present our detailed analysis of these effects for
$B\rightarrow (K^*,\;\rho, \;\omega)\gamma$ in a forthcoming paper.

We would like to acknowledge useful discussions with Dr's. Atwood, Soper, and
especially Pakvasa, Eilam and Hewett.

\newpage

\begin{figure}[htb]
\centerline{ \DESepsf(bsg.epsf width 10 cm) }
\smallskip
\caption{The branching ratio $BR$ for $b\rightarrow s\gamma$  as
a function of top quark mass with $m_b = 4.9$ GeV, $m_c = 1.5$ GeV, and
$\Lambda_4 = 0.25$ GeV. The lines 1 and 2 are for $\mu = m_b/2$ and
$\mu = 2 m_b$, respectively.
The solid (dashed) lines are for the branching ratio with (without) long
distance contribution which has been maximized by taking $a_2$ and $\kappa$
values at their top $2\sigma$ range. The allowed range is from the solid line 1
to almost the dashed line 2. }
\label{bsg}
\end{figure}

\end{document}